# Intense multi-octave supercontinuum pulses from an organic emitter covering the entire THz frequency gap

**Authors:**
C. Vicario[1], B. Monoszlai[1], M. Jazbinsek[2], S.-H. Lee[3], O-P. Kwon[3] and C. P. Hauri[1,4]

**Affiliations:**

[1]Paul Scherrer Institute, SwissFEL, 5232 Villigen PSI, Switzerland

[2]Rainbow Photonics, Zurich, Switzerland

[3]Department of Molecular Science and Technology, Ajou University, Suwon 443-749, Korea

[4]Ecole Polytechnique Fédérale de Lausanne, 1015 Lausanne, Switzerland

**In Terahertz (THz) technology, one of the long-standing challenges has been the formation of intense pulses covering the hard-to-access frequency range of 1-15 THz (so-called *THz gap*). This frequency band, lying between the electronically (<1 THz) and optically (>15 THz) accessible spectrum hosts a series of important collective modes and molecular fingerprints which cannot be fully accessed by present THz sources. While present high-energy THz sources are limited to 0.1-4 THz the accessibility to the entire THz gap with intense THz pulses would substantially broaden THz applications like live cell imaging at higher-resolution, cancer diagnosis, resonant and non-resonant control over matter and light, strong-field induced catalytic reactions, formation of field-induced transient states and contact-free detection of explosives. Here we present a new, all-in-one solution for producing and tailoring extremely powerful supercontinuum THz pulses with a stable absolute phase and covering the entire THz gap (0.1-15 THz), thus more than 7 octaves. Our method expands the scope of THz photonics to a frequency range previously inaccessible to intense sources.**

Coherent radiation in the Terahertz range (T-rays) between 0.1 and 15 THz offers outstanding opportunities in life science and fundamental research due to its non-ionizing nature. Recent progress in generating T-rays has shed a first glance onto the large potential of such radiation for controlling matter and light, such as the manipulation of Cooper pairs[1], quantum control over hydrogen atoms[2], manipulation of magnetic dynamics[3], field ionization of impurities in semiconductors[4], particle acceleration[5] and on-the-fly time lens induced by a THz pulse[6]. Different to other spectral region (near infrared, visible, x-rays) the T-rays has the potential to drive directly low-frequency motion in matter as the field oscillations are 2-4 orders of magnitude slower than for optical light and thus match the resonances of collective modes. For biology and medical applications, the non-ionizing T-rays are of growing interest for imaging and spectroscopic skin cancer diagnostics due to good spatial resolution and the ability to penetrate several millimeters of biological tissue. Present sources, however, cannot provide the required performance to exploit the full potential of the T-rays, since their emission is limited to a small frequency band between 0.1-1 THz (Lithium Niobate-based sources[7]) and 1-4 THz (organic crystal based emission[8-10]), respectively, and the spectral tunability is provided only at minor pulse energy[11]. A powerful T-ray source covering the entire THz frequency gap would be of great benefit in all the above mentioned applications for investigation and control of physical processes at frequencies, which are presently not available. Such a source would also offer substantially higher spatial resolution in biological imaging and medical applications.

Here we demonstrate a new and compact method capable to provide ultrabroadband supercontinuum T-rays at high pulse energy covering the entire THz frequency gap of 0.1 to 15 THz. The 7-octave spanning supercontinuum source outperforms modern laser and accelerator based T-ray sources, offering record-high peak fields (several MV/cm, several Tesla) and highest spectral energy density. This approach provides furthermore the capability of loss-free spectral shaping and of tuning the transform-limited pulse duration between 68fs and 1100 fs. To our knowledge this is the broadest and spectrally most intense ever produced coherent supercontinuum and the most tunable T-ray pulse source in this frequency range satisfying the requirements of above mentioned applications.

**Results**

Our original table-top Terahertz source is shown in Fig. 1. The system is based on optical rectification (OR) in the newly developed unipolar organic crystal HMQ-TMS[12] (375 µm thick) pumped by a femtosecond wavelength-tunable high-power source with output spectrum covering the near infrared spectral region of 0.8-1.5 µm. HMQ-TMS shows very promising properties for THz-wave generation since it combines large macroscopic optical nonlinearity and an exceptionally broad phase matching bandwidth due to the optimal molecular packing structure and crystal characteristics[12], which makes this crystal superior to the conventional inorganic (e.g. ZnTe, $LiNbO_3$[13]) and organic[8-10] Terahertz emitters.

**Terahertz supercontinuum generation**. Figure 2 shows the broadest THz emission extending between 0.1 and 20 THz produced by optical rectification of the 65 fs FWHM, 1.5 µm pump pulse in HMQ-TMS. The measured continuous spectrum covers more than 7 octaves with the highest spectral energy density between 2 and 10 THz. At maximum the energy content surpasses 1µJ/THz while the smallest THz energy content in the phonon-active absorption lines of the spectrum is 10 nJ/THz. The corresponding field autocorrelation of the 7-octave supercontinuum (inset Fig 2) unveils a prominent single-cycle oscillation of extremely short duration (<70 fs) with an electro-magnetic field strength of 6.1 MV/cm and 2 Tesla, respectively. The equivalence to the duration of the pump pulse is obvious and reflects the observed spectral THz cut-off at approximately 15 THz. For larger frequencies the optical rectification (also known as intra-pulse difference frequency generation) process breaks down due to the limited pump spectral width.

**Spectral pulse shaping**. For HMQ-TMS crystal the optical rectification phase-matching gives rise to a direct link between the driving laser wavelength and the produced THz spectrum. Here we show that this relation allows straightforward control of the THz spectral and temporal properties. Although pulse shaping in the frequency domain is well-established technique for visible and near-infrared lasers, spectral control is an unexplored topic for high-field pulses in the THz gap. Present strategies for shaping Terahertz pulses have been demonstrated by phase and amplitude shaping of the near-IR driving laser[14-17]. However, these shaping techniques are limited to the low frequency range (<3 THz) and to low THz peak-power.

Our approach is dramatically extending the pulse shaping capabilities in the THz gap (Fig. 3). By simply tuning the central wavelength of the pump laser between 0.8 and 1.5 µm, we could achieve phase matching condition for different THz spectral regions. Figure 3(a) shows the corresponding multi-octave spectral contents generated for different pump laser wavelengths. Ultrabroadband Terahertz radiation at a central frequency of 1 THz is, for example, produced by using the shortest pump wavelength available from our laser system ($\lambda$=800 nm). A significant spectral shift and broadening of several octaves in the THz spectrum is observed when the pump laser spectrum is tuned towards the longest available pump wavelength ($\lambda$=1500 nm). Remarkably, the produced THz spectrum covers four octaves (0.1-2 THz) for the shortest pump wavelength , and expands to more than seven octaves for λ=1500 nm. As shown in Fig. 3(b), the measured THz spectra are excellently reproduced by numerical simulations up to 12 THz, corresponding to the spectral region where the refractive index and absorption data for HMQ-TMS are known (see supplementary material section).

**Tailoring the temporal pulse shape.** The adjustable multi octave-spanning THz spectra lead to a single-cycle field transient in the temporal domain whose temporal carrier frequency varies accordingly. This behavior is measured with a THz first-order autocorrelator and is illustrated in Fig. 4 (a). The single-cycle duration of the THz field varies from 1100 fs to 68 fs for a pump wavelength being varied across the range between 0.8 and 1.5 µm. Remarkably, throughout this shaping process the THz pulse maintains its transform-limited single-cycle pulse shape. The experimentally measured field transients are accurately reproduced by numerical simulations (Fig. 4 (b)).

The spectral THz width, plotted in Fig 5(a, blue point) for each pump wavelength, shows a monotone increase for longer pump wavelengths. This results in a tunability of the spectral width between 2 and 12 THz respectively by simply varying the pump central wavelength between 800 nm and 1500 nm. Figure 5 a) (black points) illustrates also that our simple scheme allows thus the tuning of the spectral center of mass over almost 3 octaves from 1 to 6.8THz thanks to the exceptional phase-matching properties of the HMQ-TMS crystal.

The two-dimensional contour plot (Fig. 5(b)) gives an overview on the phase-matched THz frequencies and their corresponding spectral density as a function of the pump wavelength. It illustrates that phase-matching is indeed the physical explanation for the observed exceptionally large tunability in the THz spectrum for HMQ-TMS.

**Terahertz field strength.** The corresponding Terahertz transients are measured to carry an electric (magnetic) field strength as high as 6.1 MV/cm (2 Tesla) for the highest $f_C$, and a somewhat lower value for the lowest frequency (e.g. $f_C$ = 1.0THz, 100 kV/cm, 0.03 Tesla). Note that the maximum achievable field strength for a low carrier frequency is naturally smaller compared to higher frequency carrier for the same THz pulse energy which turns into a $\lambda^3$-scaling with the diffraction limited spot area and the pulse duration scaling with $\lambda^2$ and $\lambda$, respectively. In our experiment the highest field transient could potentially reach 20 MV/cm and 6.5 Tesla, respectively for sufficient pump energy to optimally pump the entire HMQ-TMS crystal. In contrast to other THz shaping schemes[14-17] the presented supercontinuum half-cycle[18] pulse synthesizer works thus at orders of magnitude higher field strengths at previously hardly accessible THz frequencies and deals without any (lossy) dispersive element on the pump laser nor on the Terahertz radiation.

In conclusion, we demonstrated for the first time a compact, laser-driven source covering the entire THz frequency gap between 0.1 and 15 THz by emitting a 7-octaves large supercontinuum at high power. The corresponding half-cycle transients are Fourier-limited and tunable in duration from 68-1100 fs at highest fields up to 6.1 MV/cm (2 Tesla). The compact scheme based on optical rectification in the organic nonlinear crystal HMQ-TMS and pumped at different optical wavelengths unravels an original pulse shaping concept which abstains lossy dispersive elements. The presented THz source attributes represents a novel and important advance for controlling matter, to modulate its electro-optical properties instantaneously, and, particularly, for investigating fundamental speed limits of ultrafast, non-resonantly field-driven processes. The record-high electro-magnetic single-cycle transients available now in the entire THz frequency gap open a new avenue to coherently steer the optical, magnetic and other properties of matter in the now accessible nonlinear regime on an ultrafast timescale, while avoiding the thermalization effects associated to a visible/mid-infrared pump source.

**Acknowledgments**

This work was carried out at the Paul Scherrer Institute and was supported by the Swiss National Science Foundation grant PP00P2_128493. CPH acknowledges association to the National Center of Competence in Research on molecular ultrafast Science and Technology (NCCR-MUST). O-P. K. and S.-H. L were supported by the National Research Foundation (NRF) grants funded by the Korean Government (MEST and MSIP) (NRF-2013K1A3A1A14055177, NRF-2013R1A2A2A01007232 and 2009-0093826). We acknowledge financial support from the Korean-Swiss Science and Technology Cooperative Program.

**Correspondence**

Correspondence and requests for materials should be addressed to C.P.H. christoph.hauri@psi.ch

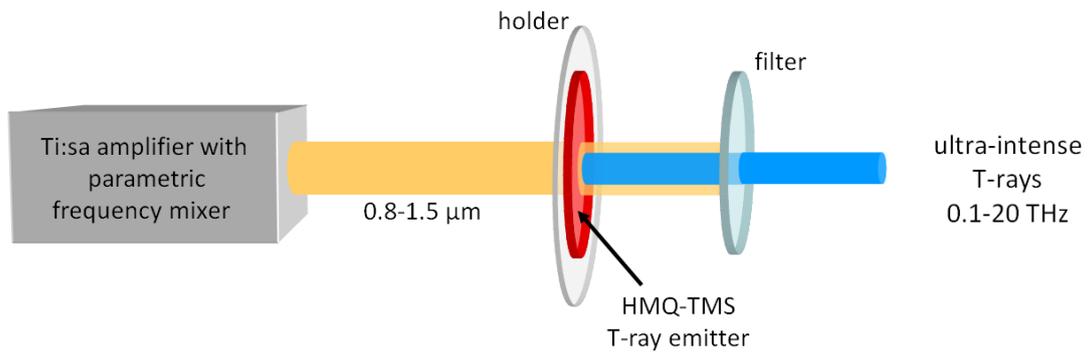

**Figure 1.** Compact laser-driven Terahertz emitter with shaping capabilities over 7 octaves. The wavelength-tunable output of an optical frequency mixer driven by a powerful Ti:sapphire laser is parametrically rectified in the nonlinear crystal HMQ-TMS, mounted on a glass substrate. Optical rectification gives rise to ultra-intense and ultra-broadband THz radiation which is separated from the pump pulse by a low-pass filter.

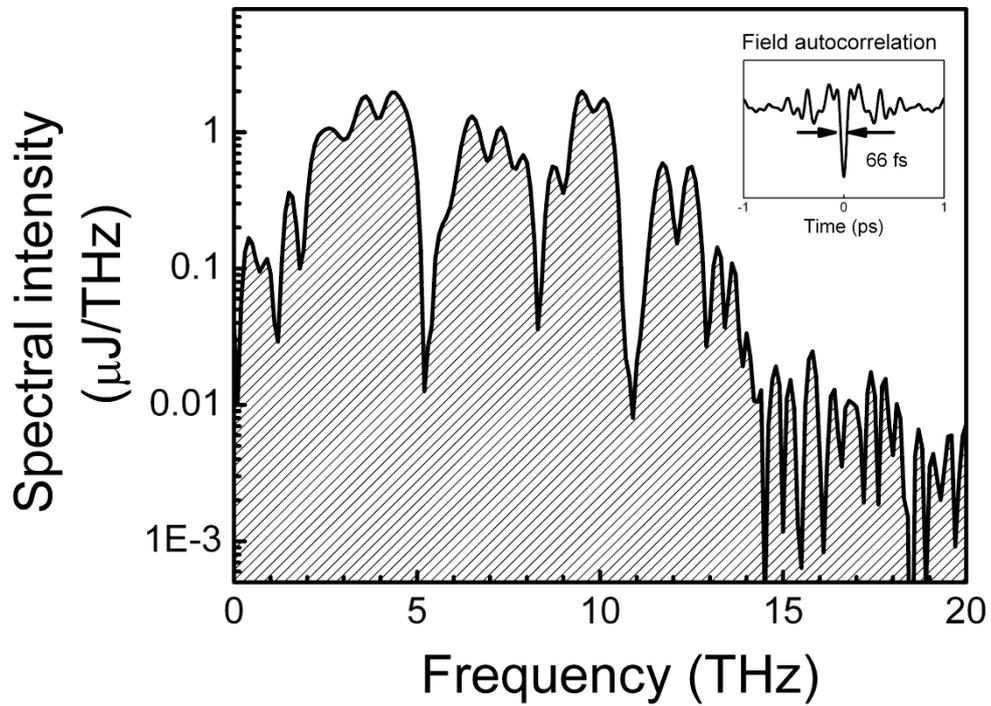

**Figure 2**

Supercontinuum Terahertz radiation produced in HMQ-TMS by optical rectification of 65 fs FWHM laser at 1500 nm central wavelength. The spectrum extends over more than 7 octaves (0.1-15 THz) and covers the entire, previously unaccessible THz gap (0.1-15 THz) at high spectral intensity. The THz field autocorrelation shown in the inset demonstrates that the spectral width is mainly limited by the pump. The phonon resonances in the HMQ-TMS are responsible for the absorption lines visible in the spectrum.

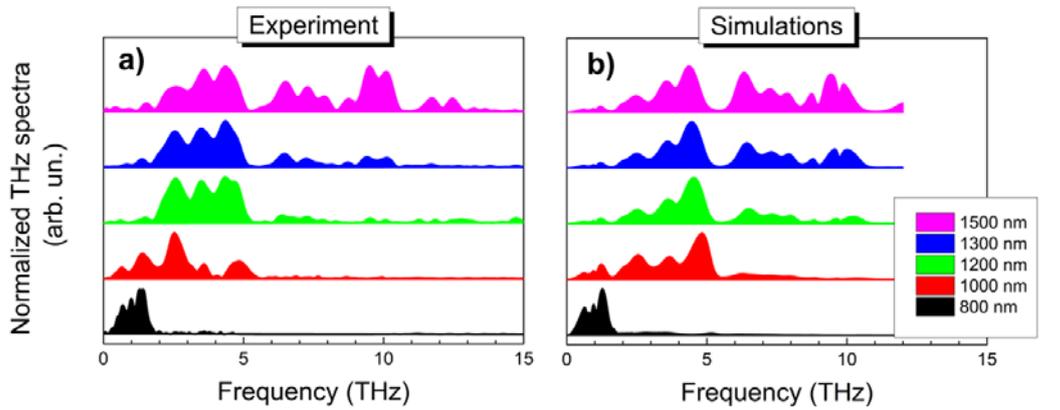

**Figure 3**

Shaping of the Terahertz spectral output by tuning the pump laser central wavelength. (a) Phase-matching conditions in HMQ-TMS give rise to a pronounced dependence of the THz spectrum as function of the pump laser wavelength which provides user control of the generated THz central frequency and the spectral width. The experimentally recorded multi-octave THz spectra cover the lower THz frequency range (800 nm pump) while the entire THz gap (0.1-15 THz) is subsequently covered for longer pump wavelength (1500 nm). The results are excellently reproduced in (b) by the corresponding simulations in the spectral range up to 12 THz (limited by the available data on refractive index).

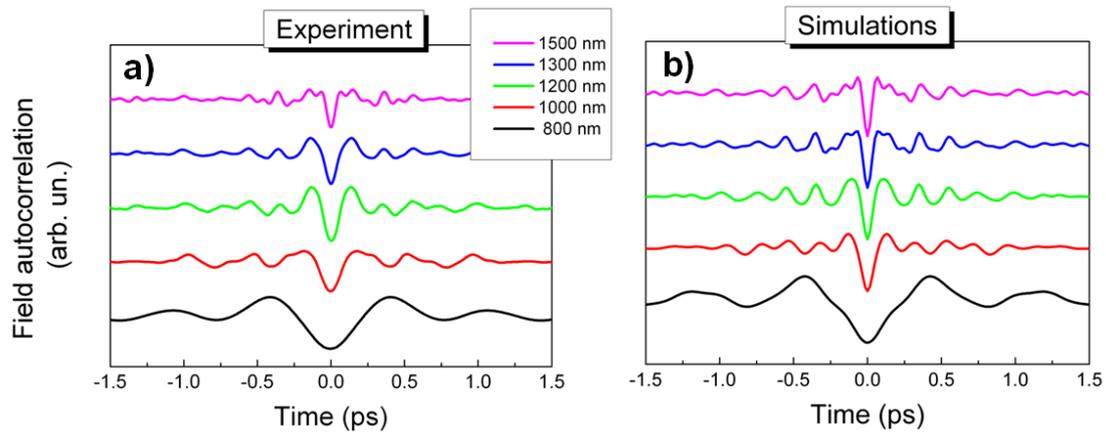

**Figure 4**

Intense half-cycle Terahertz transients produced in the T-ray emitter HMQ-TMS

(a) THz single-cycle electric field oscillations in dependence of the corresponding pump laser wavelength (same color code as in Fig. 3). The field cycle is tunable between 68 fs (purple line, 1500 nm) and 1100 fs (black line, 800 nm) while the THz half-cycle pulse characteristics are maintained. The simulated temporal field evolutions reported in b) are in excellent agreement with the experimental data.

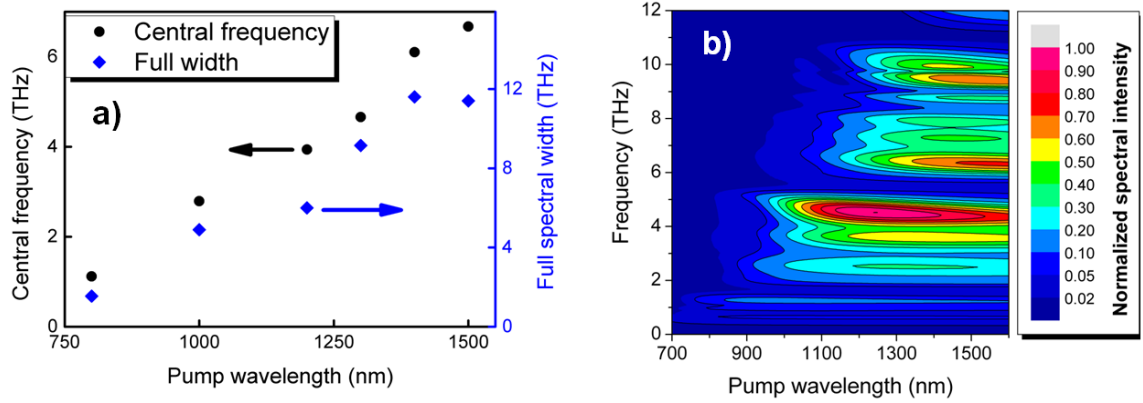

**Figure 5**

Phased-matched supercontinuum generation in HQM-TMS.

a) Central frequency (black dots) and spectral width (blue diamonds) of the THz output as a function of the pump laser central wavelength. The spectral width undergoes a dramatic extension from 2 to 12 THz by varying the pump wavelength while the centroid spectral frequency can be tuned between 1 and 6.8 THz. Shown in (b) is the calculated two dimensional phase-matching map showing the emitted THz spectral intensity as function of THz frequency and the pump wavelength for the 375 µm thick HMQ-TMS crystal used in our experiment.

**Methods**

**Optical pump**

As driving laser for the optical frequency synthesizer a 20 mJ, 50 fs FWHM pulses from a 100 Hz titanium sapphire amplifier system is used [M1]. The optical frequency synthesizer is based on optical parametric amplifiers and delivers 65 femtosecond duration FWHM, multi-mJ pulses tunable in wavelength up to 1.5 µm which are optically rectified in the highly nonlinear crystal HMQ-TMS. The small size OR crystal (7 mm in diameter and 375 µm thickness) is pumped by up to 3 mJ/cm$^2$.

**THz characterization**

For the detection of Terahertz radiation we use a Fourier-transform interferometer based on a first-order autocorrelation which has been cross-checked with an electro-optical sampling (EOS) scheme in GaP. While the EOS suffers from bandwidth-limitation associated with the detection crystal and probe pulse duration, the autocorrelation provides a flat detector response across the full THz gap (0.1-15 THz). The Terahertz pulse energy is measured by a calibrated Golay cell and the THz focal spot size is characterized by means of a micro-bolometric uncooled 2-dimensional detector (NEC). The electric field strength is calculated by measuring THz pulse energy, duration and the focal spot size.

**Simulations**

For evaluating the THz-wave generation efficiency by optical rectification in HMQ-TMS we consider the theoretical model presented in Ref [M2]. The model takes into account velocity matching between optical and THz waves, linear absorption in the optical and THz range, generation crystal thickness and pump pulse duration. The refractive indices and absorption at optical/IR frequencies of HMQ-TMS are taken from Ref [M3]. The refractive index and absorption at THz frequencies in a broad spectral range (1.2–12 THz) have been measured by THz time-domain spectrometry.

**Reference Methods**

**Supplementary material**

**Sample preparation**

The HMQ-TMS compound is synthesized by a condensation reaction, which allows a high yield and reduced impurity level compared to metathesis reaction [S1, S2]. The condensation reaction with 4-hydroxy-3-methoxybenzaldehyde and 1,2-dimethylquinolinium 2,4,6-trimethylbenzenesulfonate is performed in methanol at 70°C. HMQ-TMS crystals are grown by slow cooling method in methanol with spontaneous nucleation. As-grown HMQ-TMS crystals are cut by a knife normal to their crystallographic *b*-axis to obtain suitable facet and desired thickness for optical and THz experiments [S1].

**Refractive index and absorption of HMQ-TMS in a broad THz range**

The linear optical properties along the polar axis of HMQ-TMS crystals are measured by using THz time-domain spectroscopy (THz-TDS) in a THz spectrometer *TeraKit* from Rainbow Photonics AG, operating in a broad THz range 1.2–12 THz. The measurements are carried out using 160 µm and 450 µm thick HMQ-TMS crystals cut normal to their *b*-axis and the THz beam polarized along their polar axis. The results are shown in Fig. S1. The refractive index and the absorption data are modeled simultaneously by using a Lorentz 11-oscillator model with peaks occurring at 1.63, 2.13, 2.98, 3.92, 5.47, 6.89, 7.62, 8.34, 8.97, 9.72 and 11.02 THz. The resulting Lorentz curves are considered in the theoretical evaluation of the THz generation efficiency of HMQ-TMS crystals.

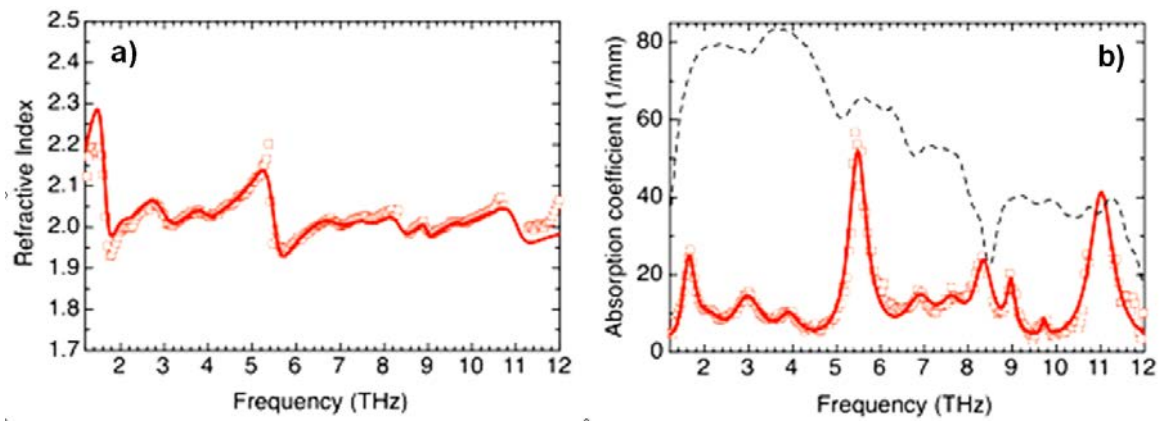

**Figure S1**. (a) Refractive index and (b) absorption coefficient along the polar axis of HMQ-TMS crystals determined by broadband THz time-domain spectroscopy. Open squares: measured data. Solid curves: best theoretical curve corresponding to a Lorentz 11-oscillator model. The dashed curve in (b) shows the dynamic range of the measurement for a 160 µm thick sample.

**Supplementary material reference**